\documentclass[12pt]{article}
\usepackage{amssymb}
\usepackage{amsmath}
\usepackage{epsfig}
\usepackage{float}
\newcommand{\be}{\begin{equation}}
\newcommand{\ee}{\end{equation}}
\newcommand{\bea}{\begin{eqnarray}}
\newcommand{\eea}{\end{eqnarray}}
\begin{document}

\begin{center}
{\bf NEUTRINO OSCILLATIONS AND TIME-ENERGY UNCERTAINTY RELATION}

\end{center}

\begin{center}
S. M. Bilenky
\end{center}

\begin{center}
{\em  Joint Institute
for Nuclear Research, Dubna, R-141980, Russia, and\\
SISSA,via Beirut 2-4, I-34014 Trieste, Italy.}
\end{center}

\begin{center}
 M. D. Mateev
\end{center}
\begin{center}
{\em  University of Sofia "St. Kliment Ohridsky", 1164 Sofia, Bulgaria}
\end{center}

\abstract{
We consider neutrino oscillations as non stationary phenomenon based on Schrodinger evolution equation 
and mixed neutrino states with definite flavor.
 We demonstrate that for such states invariance under translations in time does not take 
place.  We show that time-energy uncertainty relation plays a crucial role 
in neutrino oscillations.  We compare neutrino oscillations with $K^{0}\leftrightarrows\bar
K^{0}$, $B_{d}^{0}\leftrightarrows\bar B_{d}^{0}$ etc oscillations.}

\section{Introduction}

Evidence for neutrino oscillations
obtained in the atmospheric Super-Kamiokande \cite{SK}, solar SNO \cite{SNO},
reactor KamLAND \cite{Kamland} and other neutrino experiments 
\cite{K2K,Cl,Gallex,Sage,SKsol}
 is an  important signature of a new beyond the Standard Model Physics.

In spite of the fact that the existence of neutrino oscillations is
established, the basics of this new phenomenon is still a subject of
different opinions and active discussions (see \cite{Carlo}
and references therein). 
We will consider here neutrino oscillations 
as non stationary phenomenon based on Schrodinger evolution equation 
and notion of  mixed states for flavor neutrinos 
$\nu_{e}$, $\nu_{\mu}$ and $\nu_{\tau}$, which are produced in CC weak processes
together with, correspondingly,  $e$, $ \mu $ and $\tau $.
We will discuss flavor neutrino states in some details.
 We will show that for 
usual neutrino beams with neutrino energies many orders of magnitude larger than neutrino 
masses flavor lepton number $L_{e}$, $L_{\mu}$ and $L_{\tau}$ are effectively conserved in the 
neutrino-production and neutrino-detection SM weak processes and states of produced (and detected)
flavor neutrinos are mixed states.

The basic evolution equation of  the quantum field theory is the  Schrodinger  equation. 
According to this equation if at $t=0$ flavor neutrino (antineutrino) is produced, at the time $t$
the neutrino  (antineutrino) state is {\em non stationary one}. 
The time-energy uncertainty relation is a characteristic feature of such states
 (see, for example \cite{Messiah, Sakurai}). 
We will show that this relation plays 
a crucial role in neutrino oscillations (see \cite{SBil}). 
Neutrino oscillations, which are characterized by  finite time during which the state
of the system is significantly changed,  in accordance with time-energy uncertainty relation {\em require uncertainty in
energy}. In fact, we will demonstrate that for flavor neutrino states invariance under translations in time does not valid.

Neutrino oscillations have the same quantum-mechanical origin as 
$B_{d}\leftrightarrows \bar B_{d}$ , $K^{0}\leftrightarrows \bar K^{0}$ etc oscillations. 
We will compare here neutrino oscillations with 
$B_{d}\leftrightarrows \bar B_{d}$
oscillations which were studied recently  in details at asymmetric
$B$-factories.

\section{On the status of neutrino oscillations}
The probabilities  of the transition $\nu_{\alpha} \to \nu_{\alpha'}$ 
and $\bar\nu_{\alpha} \to \bar \nu_{\alpha'}$
 in vacuum  in the general case of $n$ neutrinos with definite masses are given by 
the following expressions (see \cite{BilPont78,BilPet87,Concha,BilGGrim99})
\begin{equation}\label{1}
{\mathrm P}(\nu_{\alpha} \to \nu_{\alpha'}) =| \sum^{n}_{i=1}
U_{\alpha' i} \,~ e^{- i\,\Delta m^2_{1i} \frac {L}{2E}} ~U_{\alpha
i}^*\, |^2.
\end{equation}
and 
\begin{equation}\label{2}
{\mathrm P}(\bar \nu_{\alpha} \to \bar \nu_{\alpha'}) =| \sum^{n}_{i=1}
U^*_{\alpha' i} \,~ e^{- i\,\Delta m^2_{1i} \frac {L}{2E}} ~U_{\alpha
i}\, |^2.
\end{equation}
Here $L$ is the distance between production and detection points,
$E$ is the neutrino energy, $\Delta m^2_{ik}=m^2_{k}-m^2_{i}$.
Indices $\alpha , \alpha' $ take the values $e, \mu, \tau, s_1, s_2,...$, indices $s_{i}$ 
label sterile neutrinos.

All existing neutrino oscillation data
(with the exception of the LSND data \cite{LSND})\footnote{Indication in favor of $\bar\nu_{\mu}\leftrightarrows\bar\nu_{e} $
 oscillations obtained several years ago in
the accelerator short-baseline  LSND experiment are going to be checked
by the running at the Fermilab
MiniBooNE experiment \cite{Miniboone}.} are 
in a good agreement with the assumption
that the number of neutrinos with definite masses 
is equal to  the number of the flavor neutrinos (three), determined from the measurement of 
the width of the decay 
of $Z^{0}$-boson into neutrino-antineutrino pairs at  the LEP experiments.
In the following we will consider the three-neutrino mixing.

The three-neutrino probabilities  of the transition $\nu_{l} \to
\nu_{l'}$ and $\bar \nu_{l} \to \bar \nu_{l'}$ in vacuum  ($l, l'= e, \mu, \tau $)   can be presented in the following form
(see, for example, \cite{BilGGrim99})
\begin{equation}\label{3}
{\mathrm P}(\nu_{l} \to \nu_{l'}) =|\delta_{l'l}+
U_{l' 2} \,~( e^{- i\,\Delta m^2_{12} \frac {L}{2E}}-1) \,~U_{l 2}^*+
U_{l' 3} \,~( e^{- i\,\Delta m^2_{13} \frac {L}{2E}}-1) \,~U_{l 3}^*
\, |^2,
\end{equation}
and 
\begin{equation}\label{4}
{\mathrm P}(\bar\nu_{l} \to \bar\nu_{l'}) =|\delta_{l'l}+
U^*_{l' 2} \,~( e^{- i\,\Delta m^2_{12} \frac {L}{2E}}-1) \,~U_{l 2}+
U^*_{l' 3} \,~( e^{- i\,\Delta m^2_{13} \frac {L}{2E}}-1) \,~U_{l 3}
\, |^2.
\end{equation}
Here
$U$ is $3\times3$  PMNS \cite{BP,MNS} mixing matrix.

In the case of the Dirac neutrinos $\nu_{i}$ the matrix $U$ is characterized by three mixing angles and
one CP phase and in the standard parametrization has the form 
\begin{eqnarray}
U=   \left (
  \begin{array}{ccc}
    c_{12} c_{13} & s_{12} c_{13} & s_{13} e^{-i \delta} \\
    -s_{12} c_{23} - c_{12} s_{23} s_{13} e^{i \delta} & c_{12} c_{23} - s_{12}
    s_{23} s_{13}e^{i \delta} & s_{23} c_{13} \\
    s_{12} s_{23} - c_{12} c_{23} s_{13}e^{i \delta} & -c_{12} s_{23} - s_{12}
    c_{23} s_{13}e^{i \delta} & c_{23} c_{13}
  \end{array}
  \right ).
\label{5}
\end{eqnarray}
Here $s_{ij} = \sin\theta_{ij}$, $c_{ij} = \cos\theta_{ij}$.

In the case of Majorana neutrinos $\nu_{i}$ the mixing matrix is given by 
\begin{equation}\label{6}
U^{M} =U\,S(\alpha),
\end{equation}
where $S_{ik}(\alpha)=e^{i\alpha_{i} }\delta_{ik}; \alpha_{3}=0$. 
Majorana phases $\alpha_{2,3}$ do not enter into the expressions (\ref{3}) and (\ref{4}) for
neutrino and antineutrino transition probabilities. Thus. investigation of neutrino oscillations can not 
allow to reveal the nature of neutrinos with definite masses  ( Majorana or Dirac?) 
\cite{BilPetHos} (see recent discussion in \cite{Bil05}).

The probabilities (\ref{3})  and (\ref{4}) depend on six parameters (two neutrino
mass-squared differences  $\Delta m^2_{12}$ and $ \Delta m^2_{23}$, 
 three mixing angles $\theta_{12}$, $\theta_{23}$ and $\theta_{13}$ and 
CP phase $\delta$) and have rather complicated
form. Taking into account the accuracy of the present-day neutrino
oscillation experiments, we can consider, however,  neutrino
oscillations in the {\em leading approximation} (see, review
\cite{BilGGrim99}).

This approximation is based on the smallness of two parameters
 \begin{equation}\label{7}
\frac {\Delta m^2_{12}}{\Delta m^2_{23}}\simeq 3.3\cdot 10^{-2};~~~
\sin^{2}\theta_{13}\leq 5\cdot 10^{-2}.
\end{equation}
The  value of the parameter $\frac {\Delta m^2_{12}}{\Delta m^2_{23}}$
can be  inferred from analysis of solar, atmospheric and KamLAND  neutrino oscillation data.
The upper bound of the parameter $ \sin^{2}\theta_{13}$ can obtained from
the data of the reactor CHOOZ experiment
\cite{Chooz}.

Let us consider first the
atmospheric-accelerator long baseline region of  $\frac{L}{E}$ in which
 \begin{equation}\label{8}
\Delta
m^2_{23}~\frac{L}{E}\gtrsim 1
 \end{equation}
In this region in the transition probabilities (\ref{2})  and  (\ref{3}) we can neglect
the contribution of the $\Delta m^2_{12}$-term. If we neglect also term
proportional to $\sin^{2}\theta_{13}$ we come to the conclusion that the only possible transitions 
in this region are
$\nu_{\mu}\to
\nu_{\tau}$ and $\bar\nu_{\mu}\to \bar\nu_{\tau}$.  For  the probabilities
of $\nu_{\mu}$ ($\bar\nu_{\mu}$) to survive
from
(\ref{2})  and  (\ref{3}) we find the standard two-neutrino expression
\begin{equation}\label{9}
{\mathrm P}(\nu_\mu \to \nu_\mu) = {\mathrm P}(\bar\nu_\mu \to
\bar\nu_\mu)= 1 - \frac {1} {2}\,\sin^{2}2\theta_{23}\, (1-\cos
\Delta m_{23}^{2}\, \frac {L} {2E}).
\end{equation}
Existing atmospheric and K2K  data are perfectly described by
(\ref{9}).
From analysis of the data of the atmospheric
Super-Kamiokande experiment the following 90 \% CL ranges of the oscillation parameters were
obtained \cite{SK}
\begin{equation}\label{10}
 1.5\cdot 10^{-3}\leq \Delta m^{2}_{23} \leq 3.4\cdot
10^{-3}\rm{eV}^{2};~~ \sin^{2}2 \theta_{23}> 0.92.
\end{equation}
For  solar and reactor KamLAND experiments small  $\Delta
m_{12}^{2}$ is relevant.
In the corresponding transition probabilities
contributions  of the ``large'' $\Delta
m_{23}^{2}$ are averaged.
For $\nu_{e}$
($\bar\nu_{e}$) survival probabilities in vacuum (or in matter) the
following general expression can be obtained  \cite{Schramm}:
\begin{equation}\label{11}
{\mathrm P}(\nu_e \to \nu_e) ={\mathrm P}(\bar\nu_e \to \bar\nu_e)=
\sin^{4} \theta_{13} +(1-\sin^{2} \theta_{13})^{2}\,{\mathrm
P}^{(12)}(\nu_e \to \nu_e),
\end{equation}
where ${\mathrm P}^{(12)}(\nu_e \to \nu_e)$ is the two-neutrino
$\nu_e$ ($\bar\nu_e$)  survival probability  in vacuum (or in matter).

If we neglect the contribution of $\sin^{2} \theta_{13}$, for the probability of reactor
$\bar\nu_{e}$ to survive in vacuum we find the following expression
\begin{equation}\label{12}
{\mathrm P}(\bar \nu_e \to \bar\nu_e)
=1-\frac{1}{2}~\sin^{2}2\,\theta_{12}~ (1 - \cos \Delta m_{12}^{2}
\,\frac {L}{2E})
\end{equation}
The probability of solar $\nu_{e}$ to survive in matter in the
approximation   $\sin^{2} \theta_{13}\to 0 $  is given by the standard two-neutrino expression which
depend on $\Delta m_{12}^{2}$, $ \tan^{2}\theta_{12}$ and electron
number density $\rho_{e}(x)$ (see, for example,  \cite{BahPenna}).

From global analysis of solar and KamLAND data it was found
\cite{SNO}
\begin{equation}\label{13}
\Delta m^{2}_{12} = 8.0^{+0.6}_{-0.4}~10^{-5}~\rm{eV}^{2};~~~
\tan^{2} \theta_{12}= 0.45^{+0.09}_{-0.07}.
\end{equation}

The existing neutrino oscillation data are compatible with two different types of neutrino mass spectra:
\begin{enumerate}
\item
normal spectrum

$m_{1}<  m_{2}    <  m_{3} ;~ \Delta m^{2}_{12}  \ll    \Delta
m^{2}_{23}  $

\item inverted spectrum\footnote{
In order to keep for the solar-KamLAND
neutrino mass-squared difference notation $\Delta m^{2}_{12}>0 $,
neutrino masses are usually labeled differently
in the cases of normal
and inverted neutrino spectra. In the case of the normal spectrum
$\Delta m^{2}_{23}>0$ and in the case of  the inverted spectrum
$\Delta m^{2}_{13}<0$.
Thus, with such notations for  the neutrino masses the character of the neutrino mass spectrum is determined by the sign
of atmospheric neutrino mass-squared difference.}
$m_{3}<  m_{1}    <  m_{2} ;~ \Delta m^{2}_{12}  \ll
 | \Delta m^{2}_{13}|  $
\end{enumerate}
In the case of the normal spectrum neutrino masses are given by
\begin{equation}\label{14}
m_{2}= \sqrt{m^{2}_{1}+\Delta m^{2}_{12} };~ m_{3}=
\sqrt{m^{2}_{1}+\Delta m^{2}_{12}+ \Delta m^{2}_{23}}
\end{equation}
For the inverted spectrum we have
\begin{equation}\label{15}
m_{1}= \sqrt{m^{2}_{3}+|\Delta m^{2}_{13}| };~ m_{2}=
\sqrt{m^{2}_{3}+|\Delta m^{2}_{13}|+ \Delta m^{2}_{12}}
\end{equation}
Neutrino mass-squared differences are known from neutrino
oscillations data. Only upper bound of the lightest neutrino mass is known at present.
From the data of the Mainz \cite{Mainz}
 and Troitsk \cite{Troitsk}
tritium $\beta$-decay experiments it was found
\begin{equation}\label{16}
m_{1(3)}\leq 2.3~ \rm{eV}.
\end{equation}
Future KATRIN \cite{Katrin} tritium experiment will be sensitive to
\begin{equation}\label{17}
m_{1(3)}\simeq 0.2 ~\rm{eV}.
\end{equation}
From cosmological data for the sum of neutrino masses upper bounds
in the range
\begin{equation}\label{18}
\sum_{i} m_{i}\leq (0.4-1.7)~ \rm{eV}
\end{equation}
can be inferred (see \cite{Tegmark}). The precision
of the cosmological measurements will significantly increase in future. 
It is expected that the future sensitivity to the sum of neutrino masses 
will reach $\sum_{i} m_{i}\simeq  0.05~ \rm{eV}$\cite{Hannestad}.

The accuracies of future neutrino oscillation 
experiments are planned to be much higher than today. 
In the experiments of the next generation one of the major efforts will be  dedicated to the measurement of the important parameter 
$\sin^{2} \theta_{13}$.  In the accelerator T2K experiment $\nu_{\mu} \to\nu_{e}  $ oscillations in the atmospheric range of the neutrino mass-squared difference will be searched for. 
The sensitivity $\sin^{2} \theta_{13}\simeq 1.5 \cdot 10^{-3}$ will be reached in this experiment \cite{T2K}. In the reactor DOUBLE CHOOZ experiment the sensitivity 
$\sin^{2} \theta_{13}\simeq 1 \cdot 10^{-2}$ is planned to be achieved \cite{Doublechooz}. If it will occur that the parameter $\sin^{2} \theta_{13}$
is not too small the  character of the neutrino mass spectrum and CP violation in the lepton sector 
can  be probed at the Super Beam\cite{Superbeam} , $\beta$- beam \cite{betabeam} and 
Neutrino Factory \cite{nufactory} facilities.

\section{Flavor neutrino states}
From the point of view of the field theory neutrino oscillations are
based on the mixing relation for the fields 
\begin{equation}\label{19}
\nu_{l L}(x) = \sum^{3}_{i=1} U_{{l}i} ~ \nu_{iL}(x)~~(l=e,\mu,\tau).
\end{equation}
 Here
$\nu_{i}(x)$ is (Majorana or Dirac) field of neutrino with mass
$m_{i}$, $U$ is the unitary PMNS mixing matrix 
and $\nu_{l L}(x)$ is so called flavor field. The flavor fields  $\nu_{l L}(x)$  enter into the standard CC and NC Lagrangians 
\begin{equation}\label{20}
\mathcal{L}_{I}^{\mathrm{CC}} = - \frac{g}{2\sqrt{2}} \,
j^{\mathrm{CC}}_{\alpha} \, W^{\alpha} + \mathrm{h.c.};~~
j^{\mathrm{CC}}_{\alpha} =2 \sum_{l=e,\mu,\tau} \bar \nu_{lL}
\gamma_{\alpha}l_{L}
\end{equation}
and
\begin{equation}\label{21}
\mathcal{L}_{I}^{\mathrm{NC}} = - \frac{g}{2\cos\theta_{W}} \,
j^{\mathrm{NC}}_{\alpha} \, Z^{\alpha};~~j ^{\mathrm{NC}}_{\alpha}
=\sum_{l=e,\mu,\tau} \bar \nu_{lL}\gamma_{\alpha}\nu_{lL},
\end{equation}
where $g$ is  the $SU(2)$ gauge constant  and $\theta_{W}$ is the weak
angle.

The relation  (\ref{19}) is the result of the diagonalization of  a neutrino mass term of the total Lagrangian.
There are two possible types of the neutrino mass terms: Majorana and Dirac 
(see \cite{BilPont78,BilPet87,Concha,BilGGrim99}).
In the case of the Majorana mass term $\nu_{i}(x)$ is the field of truly neutral Majorana particles which satisfies the condition 
\begin{equation}\label{22}
\nu_{i}(x)=\nu^{c}_{i}(x)= C~\bar\nu_{i}^{T}(x),
\end{equation}
where $C$ is the matrix of the charge conjugation.

In the case of the Dirac mass term $\nu_{i}(x)$ is the field of particles $\nu_{i}$ 
and antiparticles $\bar\nu_{i}$ which differ by the conserved total lepton number 
$L=L_{e}+L_{\mu}+L_{\tau}$ ($L(\nu_{i}) =-L(\bar\nu_{i})=1$).  

As we have mentioned before, investigation of neutrino oscillations 
does not allow to establish the nature of $\nu_{i}$. In order reveal the nature of the massive neutrinos it is necessary to study precesses in which the total lepton number $L$ is violated.
The most sensitive to the Majorana neutrino nature process is neutrinoless double $\beta$-decay
of nuclei (see \cite{betabeta})
\begin{equation}\label{23}
(A,Z) \to (A, Z+2) +e^{-} + e^-
\end{equation}

The nature of neutrinos with definite masses will be not important for our discussion. 
In this section we will consider 
the production of neutrinos (and antineutrinos) in the case of the neutrino mixing.
(see \cite{BilG01}).
Neutrinos (and antineutrinos) are produced in CC decays and reactions.
Let us consider the production of neutrinos in a CC decay
\begin{equation}\label{24}
a \to b + l^{+}+ \nu_{i},
\end{equation}
where $a$ and $b$ are some hadrons.

The 
final neutrino state is given by
\begin{equation}\label{25}
|\nu_{f}\rangle =  \sum_{i}|\nu_i\rangle  ~ \langle \nu_i,
\,l^{+}\,b\,| S |\,a\rangle,
\end{equation}
where $\langle \nu_i
\,l^{+}\,b\,| S |\,a\rangle$ is the matrix element of the process
(\ref{24}) and $|\nu_i\rangle $ is the state of left-handed neutrino with mass $m_{i}$,\footnote{Contributions  of the states with positive helicity are proportional to 
$\frac{m_{i}}{E}$ and are negligibly small}
momentum $\vec{p}$ and energy  $E_{i}= \sqrt{\vec{p}^{2}+
m^{2}_{i}}\simeq p +\frac{m^{2}_{i}}{2p}$.
We have
\begin{equation}\label{26}
H_{0}\,|\nu_i\rangle= E_{i}~|\nu_i\rangle,
\end{equation}
where $H_{0}$ is the free Hamiltonian.

In neutrino experiments energies of neutrinos $E$ are much
larger than neutrino masses: in solar and reactor experiments
$E\gtrsim 1$ MeV, in atmospheric and accelerator long-baseline
experiments  $E\gtrsim 1$ GeV etc. Taking into account that
$m_{i}\lesssim 1$ eV, we have $\frac{m^{2}_{i}}{E^{2}}\leq 10^{-12}
$. Thus, neutrino masses can be safely neglected in  matrix elements
of neutrino production processes. From
(\ref{19}) and (\ref{20}) we find
\begin{equation}\label{27}
\langle \nu_{i}\,l^{+}\,b\,| S |\,a\rangle \simeq U_{l i}^{*}\,~
\langle \nu_{l}\,l^{+}\,b\,| S |\,a\rangle_{SM},
\end{equation}
where $\langle \nu_{l}\,l^{+}\,b\,| S |\,a\rangle_{SM}$ is the
Standard Model matrix element of the process of the emission of
massless flavor neutrino $\nu_{l}$ in the decay
\begin{equation}\label{28}
a \to b + l^{+}+ \nu_{l}.
\end{equation}
We have
\begin{equation}\label{29}
\langle \nu_{l}\,l^{+}\,b\,| S
|\,a\rangle_{SM}=-i\,\frac{G_{F}}{\sqrt{2}}\,N\,2\, \bar
u_{L}(p)\,\gamma_{\alpha}\,v_{L}(p')\,\langle b|\, J^{\alpha}(0)\,|a
\rangle \,(2\pi)^{4}\, \delta (P'-P).
\end{equation}
Here $N$ is the product of the standard normalization factors, $p$ is neutrino momentum, 
$p'$
is the momentum of $l^{+}$,  $P$ and $P'$
are total initial and final momenta and $J^{\alpha}$ is hadronic
charged current.\footnote{ The 
arguments presented above are not applicable to high-energy part of $\beta$-spectrum
of the decay $^{3}\rm{H}\to ^{3}\rm{He}+e^{-}+\bar\nu_{e}$ which
corresponds to the emission of neutrino with {\em energy comparable
with neutrino mass}. For the spectrum we have \bea
\frac{d\,\Gamma}{d\,E} =C\,p\,(E+m_{e})\,(E_{0}-E)\,
\sum_{i}|U_{ei}|^{2}~\sqrt{(E_{0}-E)^{2}-m_{i}^{2}}\,F(E)\,\theta(E_{0}-E-m_{i}),
\nonumber\eea where $E_{0}$ is the energy, released in the decay, $
m_{e}$ is the mass of the electron, $F(E)$  is the Fermi function
which takes into account the Coulomb interaction of the final
particles and $C$  is a constant. From the measurement of the electron
spectrum in the high-energy region the bound (\ref{16}) 
 was obtained in the Troitsk and the
Mainz experiments.}

From (\ref{25}) and (\ref{27}) for the final neutrino 
state we find the expression
\begin{equation}\label{30}
|\nu_{f}\rangle =  |\nu_{l}\rangle ~\langle \nu_{l}\,l^{+}\,b\,| S
|\,a\rangle_{SM},
\end{equation}
where
\begin{equation}\label{31}
|\nu_{l}\rangle=\sum^{3}_{i=1}U^{*}_{li}~|\nu_{i}\rangle.
\end{equation}
is normalized left-handed neutrino state.  It follows from (\ref{30})
that the probability  of the decay (\ref{28}) is given by the Standard Model (assuming  that 
$\nu_{l}$  is massless).

Neutrino which is produced in a CC weak decay together with $l^{+}$ is called
flavor neutrino
$\nu_{l}$. We have shown that the state of flavor
neutrino is given by {\em coherent superposition} of
states of neutrinos with definite masses.

 Analogously, in CC processes together with lepton
$l^{-}$ right-handed flavor antineutrino $\bar\nu_{l}$ is produced.
The state of  $\bar\nu_{l}$ is given by the expression
\begin{equation}\label{32}
|\bar\nu_{l}\rangle = \sum_{i=1}^{3} U_{l i}~ |\nu_i\rangle,
\end{equation}
where $|\nu_i \rangle$ is the state of the right-handed neutrino (or right-handed antineutrino in the Dirac case)
 with mass $m_{i}$, momentum
$\vec{p}$ and energy $E_{i}\simeq p +\frac{m^{2}_{i}}{2p}$.

We will consider now detection of neutrinos with energies much larger than neutrino masses.
Neutrinos are detected via the observation of CC and NC weak
processes.
Let us consider,  for example, inclusive process 
\begin{equation}\label{33}
\nu_{l'}+N\to l^{-} + X .
\end{equation}
Neglecting neutrino masses,
for the matrix element of the process we have
\begin{eqnarray}
\langle l\,X\,|~ S~ |\,\nu_{l'}~N\rangle&=& \sum_{i}\langle
l\,X\,|~ S~ |\,\nu_{i}~N\rangle ~U^{*}_{l' i}\nonumber\\
=\langle l\,X\,|~ S~ |\,\nu_{l}~N\rangle
_{SM}~\sum_{i}U^{*}_{l'i}~U_{li} &=&\langle l\,X\,|~ S~
|\,\nu_{l}~N\rangle _{SM}~\delta_{l'l}\label{34},
\end{eqnarray}
where
\begin{equation}\label{35}
\langle l\,X\,|~ S~ |\,\nu_{l}\,N\, \rangle_{SM}=
-i\,\frac{G_{F}}{\sqrt{2}}\,N\,2\, \bar
u_{L}(p')\,\gamma_{\alpha}\,u_{L}(p)\,\langle X|\, J^{\alpha}(0)\,|N
\rangle \,(2\pi)^{4}\, \delta (P'-P),
\end{equation}
where $p'$ is the momentum of final lepton and $p$ is the neutrino momentum.

It follows from (\ref{34})  that due to unitarity of the neutrino mixing matrix
the matrix element $\langle l\,X\,|~ S~ |\,\nu_{l'}~N\rangle$
is different from zero only if $l'=l.$. Thus,  the lepton $l^{-}$ can be produced 
in CC process (\ref{33}) 
by the left-handed flavor neutrino $\nu_{l}$.
Analogously, the lepton $l^{+}$ can be produced in inclusive CC process 
\begin{equation}\label{36}
\bar \nu_{l}+N\to l^{+} + X .
\end{equation}
by the right-handed flavor antineutrino $\bar \nu_{l}$.
We come to the conclusion that  in CC  processes flavor lepton numbers are effectively 
conserved.

Let us summarize previous discussion.
For neutrinos with energies many orders of magnitude larger than neutrino masses in matrix elements of neutrino-production and neutrino-detection processes
neutrino masses can be neglected. As a result of that
\begin{itemize}
\item
Lepton flavor numbers
$L_{e}$, $L_{\mu}$ and  $L_{\tau}$ are conserved in such processes: together with
$l^{-}$ right-handed flavor antineutrino $\bar \nu_{l}$ is produced, flavor  left-handed
neutrinos  $\nu_{l}$ in the processes of interaction with nucleon produce $l^{-}$ etc.

Non conservation of the flavor lepton numbers can be revealed only
in such processes in which effects of neutrino masses are relevant.
Such processes are neutrino oscillations in vacuum and neutrino
transitions in matter.

\item
Matrix elements of  neutrino-production and neutrino-detection processes  are given by the Standard Model expressions
(in which neutrino masses  are neglected)
\item

States of flavor neutrino $\nu_{l}$ and antineutrinos $\bar \nu_{l}$
are given by coherent superpositions  (\ref {31}) and (\ref {32}).

\end{itemize}
Let us stress that states of flavor neutrino $\nu_{l}$ and flavor antineutrino $\bar\nu_{l}$ are  the  superpositions 
of the states  of neutrinos with definite masses $\nu_{i}$ with coefficients $U^{*}_{li}$  and $U_{li}$, respectively.
Because of this difference in the case of the  CP
violation 
$$P (\nu_{l}\to \nu_{l'})\neq P (\bar\nu_{l}\to \bar\nu_{l'});~~ l' \neq l.$$

The mixed flavor neutrinos and antineutrinos states  (\ref {31})  and  (\ref {32}) 
are different from
usual states of particles in the Quantum Field Theory. We will show now that for such state
invariance under translation in time is not valid.

Let us consider translations in space and time (see, for example \cite{BogShir})
\be
x'_{\alpha} = x_{\alpha} +a_{\alpha},
\label{37}
\ee
where $a$ is a constant vector. 
In the case of the invariance under translation we have
\be
|\Psi \rangle' = e^{i\,P\, a }\, |\Psi \rangle,
\label{38}
\ee
and
\be
O(x+a)=  e^{i\,P\, a }\,O(x)\, e^{-i\,P\, a },
\label{39}
\ee
where $P_{\alpha}$ is the operator of the total momentum, $O(x)$ is any  operator and 
vectors $|\Psi \rangle $ and $|\Psi \rangle'$ describe  {\em the same} physical state.

If $|\Psi \rangle $ is a state with total momentum
$p$ vectors $|\Psi \rangle$ and $|\Psi \rangle'$ differ by the phase factor
\be
|\Psi \rangle' = e^{i\,p\, a }\, |\Psi \rangle .
\label{40}
\ee
Let us apply now the operator of the translations $e^{i\,P\,a}$ to the mixed flavor neutrino state
$|\nu_{l}\rangle$ given by Eq. (\ref{31}).
We have
\be
|\nu_{l}\rangle'=    e^{i\,P\, a }\, |\nu_{l}\rangle =  e^{-i\,\vec{p}\,\vec{a}}\,\sum_{l'}
|\nu_{l'}\rangle \,\sum_{i}U_{l' i}e^{i\,E_{i}\,a^{0}}\,U_{l i}^*.
\label{41}
\ee
The  vector $|\nu_{l}\rangle'$ describes
 {\em superposition of different flavor states}. Thus,
initial and transformed vectors describe {\em different states}.
We come to the conclusion that in the case of the states
which describe mixed flavor neutrinos with definite momentum $\vec{p}$ there is no invariance under translation in time. This
means that in   transitions between different flavor neutrinos (and antineutrinos) energy is not conserved.

\section{Neutrino oscillations in vacuum and time-energy uncertainty relation}
The basic  evolution equation of the quantum field theory is the Schrodinger
equation
\begin{equation}\label{42}
i\,\frac{\partial |\Psi(t) \rangle}{\partial t} = H\, |\Psi(t)
\rangle,
\end{equation}
where $ H$ is the total Hamiltonian. The general solution of the
equation (\ref{42}) has the form
\begin{equation}\label{43}
 |\Psi(t) \rangle=e^{-i Ht}~ |\Psi(0) \rangle,
\end{equation}
where $|\Psi(0) \rangle$ is an initial state.

Let us consider the evolution in vacuum of the  states of flavor
neutrinos $\nu_{l}$  and flavor antineutrinos 
$\bar \nu_{l}$ 
produced in weak processes. We have in this case
\begin{equation}\label{44}
|\Psi(0) \rangle=|\nu_{l}\rangle;~~~\rm{or}  ~~ ~ |\Psi(0) \rangle=|\bar \nu_{l}\rangle ; ~~~H=H_{0},
\end{equation}
where $H_{0}$ is the free Hamiltonian and the states
$|\nu_{l}\rangle$ and $|\bar\nu_{l}\rangle$ are  given by Eq.(\ref{31}) 
and   Eq.(\ref{32}). Now, taking into
account (\ref{43})
 for neutrino and antineutrino states at the
time $t\geq 0$ we have
\begin{equation}\label{45}
|\nu_{l}\rangle_{t} = \sum_{i=1}^{3}|\nu_i\rangle\,
e^{-iE_{i}\,t}\,U_{l i}^*
\end{equation}
and
\begin{equation}\label{46}
|\bar \nu_{l}\rangle_{t} = \sum_{i=1}^{3}|\nu_i\rangle\,
e^{-iE_{i}\,t}\,U_{l i}
\end{equation}

Thus,  flavor
neutrinos $\nu_{l}$ and antineutrinos $\bar\nu_{l}$,  produced in weak
processes at $t=0$,  at $t>0$ are  described by {\em non stationary
states}.

It is a general property of quantum theory that for non stationary
states the time-energy uncertainty relation
\begin{equation}\label{47}
\Delta E~\Delta t \geq 1
\end{equation}
takes place (see, for example,  \cite{Messiah,Sakurai,Bauer}). In this relation
$\Delta E $ is uncertainty in energy and $\Delta t$ is time interval
during which significant changes in the
 system happen.

In the neutrino case  
\begin{equation}\label{48}
(\Delta E)_{ik}=E_{k}-E_{i}\simeq \frac{\Delta m^{2}_{ik}}{2E}
\end{equation}
and time-energy uncertainty relation takes the form
\begin{equation}\label{49}
\frac{\Delta m^{2}_{ik}}{2E}~t \geq 1.
\end{equation}
where $t$ is the time interval during which flavor content of
neutrino state is significantly changed.

Neutrinos are detected through the   observation  of weak processes. 
Let us develop the state $|\nu_{l}\rangle_{t}$ over
flavor states $|\nu_{l'}\rangle$. We have
\begin{equation}\label{50}
|\nu_{l}\rangle_{t} =\sum_{l'}|\nu_{l'}\rangle~\mathbf{A}(\nu_{l}\to
\nu_{l'};t),
\end{equation}
where
\begin{equation}\label{51}
\mathbf{A}(\nu_{l}\to \nu_{l'};t)= \sum_{i=1}^{3}U_{l'i}~
e^{-iE_{i}\,t}\,U_{l i}^*
\end{equation}
is the amplitude of the transition $\nu_{l}\to \nu_{l'}$ during the
time $t$.\footnote{We  can obtain the same result in another way. Let us consider the process
$\nu_{i}+N\to l' + X $. For neutrinos with energies many order of magnitude larger than neutrino masses we have 
$$\langle l\,X\,|~ S~ |\,\nu_{i}~N\rangle \simeq \langle l'\,X\,|~ S~ |\,\nu_{l'}~N\rangle_{SM}~U_{l'i}$$
From (\ref{45}) and this relation we will find the expression (\ref{51}) for 
$\nu_{l}\to \nu_{l'}$ transition amplitude.}

 Analogously in the case of the antineutrino we have
\begin{equation}\label{52}
|\bar\nu_{l}\rangle_{t} =\sum_{l'}|\bar
\nu_{l'}\rangle~\mathbf{A}(\bar\nu_{l}\to \bar\nu_{l'};t),
\end{equation}
where the amplitude of the transition $\bar\nu_{l}\to \bar\nu_{l'}$
is given by
\begin{equation}\label{53}
\mathbf{A}(\bar\nu_{l}\to \bar\nu_{l'};t)= \sum_{i=1}^{3}U^*_{l'i}~
e^{-iE_{i}\,t}\,U_{l i}
\end{equation}
The expressions (\ref{51}) and  (\ref{53}) have a simple meaning: 
$U_{l i}^*(U_{l i})$ is the amplitude of the transition from initial flavor neutrino (antineutrino) state  $|\nu_{l}\rangle  (|\bar\nu_{l}\rangle)$ to the state $|\nu_{i}\rangle$; the factor 
$e^{-iE_{i}t}$ describes propagation in the state with definite energy $E_{i}$ and 
$U_{l' i}(U^*_{l' i})$ is the amplitude of the transition from the state  $|\nu_{i}\rangle$ to the final flavor state $|\nu_{l'}\rangle  (|\bar\nu_{l'}\rangle)$. 
Because neutrino masses can not be resolved in the production and detection processes
in the amplitudes 
 (\ref{51}) and  (\ref{53})  sum over all $i$ is performed. 

It is instructive to derive expression  (\ref{51}) and  (\ref{54}) starting from the flavor representation.
We have 
\begin{equation}\label{54}
|\Psi(t)\rangle=\sum_{l=e,\mu,\tau}|\nu_{l}\rangle~a_{l}(t),
\end{equation}
where $~a_{l}(t)= \langle \nu_{l}|\Psi(t)\rangle$ is the wave function of neutrino in the flavor
 representation. From (\ref{42} )  for the equation of motion in vacuum we find 
\begin{equation}\label{55}
 i\,\frac{\partial\, a(t)}{\partial t}=\mathrm{H}_{0}\,a(t), 
\end{equation}
where 
\begin{equation}\label{56}
(\mathrm{H_{0}})_{l'l}   =\langle \nu_{l'}|H_{0}|\nu_{l}\rangle = \sum_{i}
U_{l'i}~E_{i}~U^{*}_{li}= (U~E~U^{\dagger})_{l'l}
\end{equation}
is the free Hamiltonian in the flavor representation. In order to obtain the solution of the equation 
(\ref{55}) let us introduce the function
\begin{equation}\label{57}
a'(t) = U^{\dagger}~a(t).
\end{equation}
From (\ref{55})  and (\ref{57}) we have  
\begin{equation}\label{58}
i\,\frac{\partial\, a'(t)}{\partial t}=E\,a'(t).
\end{equation}
The solution of this equation is obvious:
\begin{equation}\label{59}
a'(t) =e^{-i Et}~a'(0) 
\end{equation}
From (\ref{57} )  and (\ref{59}) we find that solution of the equation (\ref{55}) is given by
\begin{equation}\label{60}
a(t) =U~e^{-i Et}~ U^{\dagger} ~a(0)
\end{equation}

Assuming that $a_{l''}(0) =\delta_{l''l}$ for 
the amplitude of the transition $\nu_{l}\to\nu_{l'} $ we find the expression
\begin{equation}\label{61}
a_{l'}(t)=(U~e^{-i Et}~ U^{\dagger})_{l'l},
\end{equation}
which  coincides with (\ref{51}).

Analogously, in the case of antineutrino we have
\begin{equation}\label{62}
|\Psi(t)\rangle=\sum_{l=e,\mu,\tau}|\bar \nu_{l}\rangle~b_{l}(t),
\end{equation}
where $~b_{l}(t)= \langle \bar\nu_{l}|\Psi(t)\rangle$ is the wave function of antineutrino in the flavor
 representation. The function $b(t)$ satisfies the following evolution equation
\begin{equation}\label{63}
 i\,\frac{\partial\, b(t)}{\partial t}=\mathrm{\bar H}_{0}\,b(t), 
\end{equation}
where 
\begin{equation}\label{64}
(\mathrm{\bar H_{0}})_{l'l}   =\langle \bar \nu_{l'}|H_{0}|\bar \nu_{l}\rangle 
= (U^{*}~E~U^{T})_{l'l}
\end{equation}
is the free Hamiltonian in the flavor representation. The solution of this equation is given by
\begin{equation}\label{65}
b(t) =U^{*}~e^{-i Et}~ U^{T} ~b(0).
\end{equation}
If we assume  that $b_{l''}(0) =\delta_{l''l}$ for 
the amplitude of the transition $\bar\nu_{l}\to \bar\nu_{l'} $ we find the expression
\begin{equation}\label{66}
b_{l'}(t)=(U^{*}~e^{-i Et}~ U^{T})_{l'l}.
\end{equation}
which coincides  with (\ref{53}).

From  (\ref{51}) and  (\ref{53}) for the probabilities of the transitions  
 $\nu_{l}\to \nu_{l'}$ and  $\bar\nu_{l}\to \bar\nu_{l'}$ in vacuum we obtain the following 
standard expressions
\begin{equation}\label{67}
P(\nu_{l} \to\nu_{l'}) = |\mathbf{A}(\nu_{l}\to \nu_{l'};t)|^{2}=
|\delta_{l' l}+ \sum_{i=2,3} U_{l' i}
\,(e^{-i \Delta  m^2_{ 1i }\frac{L }{2 E }} -1) U^{*}_{li}|^{2}
\end{equation}
and 
\begin{equation}\label{68}
P(\bar\nu_{l} \to\bar \nu_{l'}) = 
|\mathbf{A}(\bar\nu_{l}\to \bar\nu_{l'};t)|^{2}=
|\delta_{l' l}+ \sum_{i=2,3}
U^{*}_{l' i} \,(e^{-i \Delta  m^2_{ 1i }\frac{L }{2 E }} -1)
U_{li}|^{2}.
\end{equation}
Taking into account the unitarity of the neutrino mixing matrix it is easy to check that 
$P(\nu_{l} \to\nu_{l'})$ and $P(\nu_{l} \to\nu_{l'})$ are normalized probabilities:\footnote{We assumed that states $|\nu_{i}\rangle$ 
are the states
with the same momentum $\vec{p}$. Correspondingly, mixed states $|\nu_{l}\rangle$ and 
$|\bar \nu_{i}\rangle$ are characterized by momentum $\vec{p}$. 
Notice that this is the standard procedure,  which corresponds 
to conditions of experiments with beams of particles. If, we assume, however, that 
$|\nu_{i}\rangle$ are states  
 with different momenta $\vec{p_{i}}$, for oscillation phases in (\ref{67}) and  (\ref{68})
we will have 
$(E_{i}- E_{1})t\simeq [(p_{i}-  p_{1})L +  \Delta  m^2_{ 1i }\frac{L}{ 2 E }] $ , where the first term
 $p_{i}-  p_{i}$ is proportional to  $\Delta m^2_{ 1i }$ with some unknown coefficient which could vary from one experiment  to another. There is no such term in the oscillation phases:
data of different neutrino oscillation experiments 
are compatible if the standard expressions 
(\ref{67}) and  (\ref{68}) for transition probabilities are used. 
In the framework of the considered formalism with mixed flavor neutrino states and 
Schroedinger evolution equation ``the equal momentum assumption''  is the only 
possibility to obtain  oscillation phases  which do not depend on experimental conditions     ''}
\begin{equation}\label{69}
\sum_{l'=e,\mu,\tau} P(\nu_{l} \to\nu_{l'}) =1;~~
\sum_{l'=e,\mu,\tau}P(\bar\nu_{l} \to\bar \nu_{l'})=1
\end{equation}
In (\ref{67}) and  (\ref{68}) we have used the relation 
\begin{equation}\label{70}
t\simeq L,
\end{equation}
where $L$ is the distance between neutrino-production and
neutrino-detection points. There were many discussions in literature connected with  the
relation (\ref{70}) (see \cite{Carlo, Kayser}). After K2K experiment \cite{K2K}
there is no reasons for such discussions. In this experiment this
relation was confirmed and used to select neutrino events.

In the K2K experiment neutrinos were produced by protons from the
KEK accelerator in  1.1 $\mu sec$ spills.  Protons were extracted
from the accelerator every 2.2 sec. The difference of the time of the
detection of neutrinos in the Super-Kamiokande detector ($t_{SK}$)
and the time of the production of neutrinos at KEK ($t_{SK}$)  $t=
t_{SK}-t_{KEK}$ was measured in K2K experiment. Let us determine
\begin{equation}\label{71}
\Delta t\simeq t-t_{TOF},
\end{equation}
where $t_{TOF}=L/c$.  In the K2K experiment muon neutrino events
which satisfy the criteria
$$-0.2\leq\Delta t \leq 1.3 \, \mu sec$$
were selected. Notice 
that in the K2K experiment  $L \simeq 250$ km and $ L/c\simeq 0.83\cdot 10^{3}\,\mu
sec$.

It follows from (\ref{67}) and (\ref{68}) that neutrino oscillations
can be observed if at least  for one value of $i$ the following
inequality is satisfied (see \cite{BilPont78,BilPet87})\footnote{It is obvious that
inequality (\ref{72}) is only necessary condition of the observation
of neutrino oscillations. For the transition $\nu_{l} \to \nu_{l'}$
to be observed, corresponding elements of the neutrino
 mixing matrix have to be not small.}
\begin{equation}\label{72}
\frac{\Delta m^{2}_{1i}}{2E}~L \geq 1
\end{equation}
Comparing (\ref{49}) and (\ref{72}) we conclude that in the case of
neutrino oscillations time-energy uncertainty relation coincides
with  the condition of the observation of neutrino oscillations\cite{SBil}.

Let us stress that finite time  during which a significant change of
the flavor content of neutrino state happens (oscillation time) in
accordance with time-energy uncertainty relation requires
uncertainty in energy. This corresponds to the violation of the
invariance under translation in time in the case of the mixed
neutrino states, which we discussed in the previous section.

We will make the following remark. It was  stated
in some papers  (see \cite{Kayser, Stodol,Lipkin}) that neutrino oscillations can
take place only if energies of different neutrinos $\nu_{i}$ are
equal. This statement is based on the assumption that in experiments
on the study of neutrino oscillations 
only distance $L$ is relevant. The 
time is not measured and
transition probability must be averaged over time. 

We do not see any reasons for such  assumption.
Of course, the time of the traveling of neutrinos from production to detection points is not 
measured in the solar, atmospheric and reactor experiments. However, from 
K2K  experiment, in which this time is measured,   we know that traveling time and distance between production and detection points 
are equal.

We will present another argument against "equal energy"
assumption. Let us consider the propagation of neutrino in matter. The 
evolution equation in the flavor representation has the form
\begin{equation}\label{73}
i\,\frac{\partial a(t)}{\partial t}= \mathrm{H}\,a(t).
\end{equation}
Here
\begin{equation}\label{74}
\mathrm{H}= \mathrm{H_{0}}+\mathrm{H_{I}},
\end{equation}
where $\mathrm{H_{0}}$ is the free Hamiltonian and 
$\mathrm{H_{I}}$ the effective Hamiltonian of interaction of neutrino with matter. 

The free Hamiltonian in the flavor representation is
given by
\begin{equation}\label{75}
(\mathrm{H_{0}})_{l'l}   =\langle \nu_{l'}|H_{0}|\nu_{l}\rangle = \sum_{i}
U_{l'i}~E_{i}~U^{*}_{li}.
\end{equation}

The refraction indices of flavor neutrinos in matter are determined by amplitudes of 
elastic neutrino scattering in forward direction and target particles densities. Taking into account 
$\nu_{e}-\nu_{\mu}-\nu_{\tau}$ universality of NC  for the effective Hamiltonian of interaction 
of neutrino with matter we have \cite{Wolf}
\begin{equation}\label{76}
(\mathrm{H_{I}})_{l'l}=\sqrt{2}\,G_{F}\,\rho_{e}\,\eta_{l'l},
\end{equation}
where $\rho_{e}$ is electron number density, $\eta_{ee}=1$, other elements
of $\eta_{l'l}$ are equal to zero. Let us stress that $\mathrm{H_{I}}$ is determined by the CC part 
of the Standard Model amplitude of the $\nu_{e}e\to\nu_{e}e $ forward scattering. Neutrino masses 
 in the interaction Hamiltonian do not enter. 

If  energies of neutrinos with definite masses are equal
($E_{i}=E$), in this case  the free Hamiltonian is unit matrix 
\begin{equation}\label{77}
(\mathrm{H_{0}})_{l'l}
=E~\delta_{l'l}
\end{equation}
 With such free Hamiltonian it will be no matter effect
\cite{MSW} which was  observed in solar neutrino experiments
\cite{Lisi}. Thus, assumption of "equal energies" is not compatible with
data of solar neutrino experiments.

\section{Comparison of neutrino oscillations with
$B^{0}_{d}\leftrightarrows \bar
B^{0}_{d}$  oscillations}

Neutrino oscillations and flavor oscillations of neutral mesons
($K^{0}\leftrightarrows \bar K^{0}$, $B^{0}_{d}\leftrightarrows \bar
B^{0}_{d}$ etc) have the same
quantum-mechanical origin.\footnote{ In fact, the existence of 
$K^{0}\leftrightarrows \bar K^{0}$ oscillations was 
major argument for B. Pontecorvo \cite{BP} to propose
 neutrino oscillations in 1957.}

We will compare here
neutrino oscillations with  $B_{d}\leftrightarrows \bar B_{d}$
oscillations which were studied recently in details at asymmetric
$B$-factories \cite{bfactories}.

The states of $B_{d}$ and $\bar B_{d}$ mesons are eigenstates of the
Hamiltonian $H_{0}$ which  is the  sum of
the free Hamiltonian and Hamiltonians of the strong and
electromagnetic interactions. Assuming  CPT invariance of the strong
interaction in the rest frame of  ($B_{d}^{0} (  \bar B_{d}^{0})$  we have
\begin{equation}\label{78}
H_{0}~|B_{d}^{0}\rangle=m_{B}~|B_{d}^{0}\rangle;~~ H_{0}~|\bar
B_{d}^{0}\rangle=m_{B}~|\bar B_{d}^{0}\rangle,
\end{equation}
where $m_{B}$ is the mass of $B_{d}$ ($\bar B_{d}$) meson.

$B^{0}_{d}$
and $\bar B^{0}_{d}$ mesons are produced in the decays of $\Upsilon
(4S)$ and other strong processes in which quark flavor is conserved.
Effects of weak interaction can be neglected in production processes.
After $B^{0}_{d}$ and $\bar B^{0}_{d}$ mesons are produced
weak interaction plays the major role: due to weak interaction
particles decay, eigenstates of the total effective Hamiltonian have
different masses  etc.

Let us assume that at $t=0$ $B^{0}_{d}$ ($\bar B^{0}_{d}$) was produced. At  $t>0$
for the vector of the state we have 
\begin{equation}\label{79}
|\Psi (t)\rangle= \sum_{\alpha=B_{d},\bar B_{d}}  a_{\alpha}(t)~ |  \alpha  \rangle + 
\sum_{i}b_{i}(t)~| i  \rangle,
\end{equation}
where $a_{\alpha}(t)$ is the wave function of $B^{0}_{d}-\bar B^{0}_{d}$ system in the 
flavor representation and the states $| i  \rangle$ describe products of the decay of 
$B^{0}_{d}$ and $\bar B^{0}_{d}$ . The   vector $|\Psi (t)\rangle$  satisfies the Schrodinger 
equation (\ref{42}). From  (\ref{79}) and  (\ref{42}) it can be shown (see \cite {echaja}) that in  the Waiskopf-Wigner approximation
the function $a(t)$ satisfies the evolution equation 
\begin{equation}\label{80}
i\,\frac{\partial\, a(t)}{\partial t}=\mathrm{H}\,a(t).
\end{equation}
Here $\mathrm{H}$ is the total effective non hermitian Hamiltonian
of $B^{0}_{d}-\bar B^{0}_{d}$ system in the flavor representation. The Hamiltonian $\mathrm{H}$ has
the form
\begin{equation}\label{81}
\mathrm{H}= \mathrm{M} -\frac{i}{2}~\mathrm{\Gamma},
\end{equation}
where $\mathrm{M}=\mathrm{M}^{\dag}$ and $\mathrm{\Gamma}=\mathrm{\Gamma}^{\dag}$ are $2\times2 $
matrices of mass  and width.

For the eigenstates of the total Hamiltonian we have
\begin{equation}\label{82}
\mathrm{H}~a_{H}=\lambda_{H}~a_{H};~~~\mathrm{ H}~a_{L}=\lambda_{L}~a_{L}.
\end{equation}
Here
\begin{equation}\label{83}
\lambda_{H}=m_{H}-\frac{i}{2}~\Gamma_{H};~~~
\lambda_{L}=m_{L}-\frac{i}{2}~\Gamma_{L},
\end{equation}
where $m_{H}$,  $m_{L}$ and $\Gamma_{H}$, $\Gamma_{L}$ are masses
and total widths of $B_{H}^{0}$ and $B_{L}^{0} $  mesons.

From (\ref{82}) for the states of $B^{0}_{H}$ and $B^{0}_{L}$ mesons
we have
\begin{equation}\label{84}
|B_{H}^{0}\rangle=N~(|B_{d}^{0}\rangle+\frac{q}{p}~|\bar
B_{d}^{0}\rangle);~~|B_{L}^{0}\rangle=N~(|B_{d}^{0}\rangle-\frac{q}{p}~|\bar
B_{d}^{0}\rangle ).
\end{equation}
Here $N=1/\sqrt{1+|q/p|^{2}}$, $q=\sqrt{H_{\bar B_{d}~B_{d}}}$ and
$p=\sqrt{H_{ B_{d}~\bar B_{d}}}$. The parameter $\frac{q}{p}$
characterize CP violation of the weak interaction (if CP is
conserved $\frac{q}{p}=1$).

For the states of $B_{d}^{0}$ and $\bar B_{d}^{0}$ mesons, produced in 
the decays of $\Upsilon (4S)$ and in other strong processes, from (\ref{84})   we have
\begin{equation}\label{85}
|B^{0}_{d}\rangle=\frac{1}{2N}~(|B^{0}_{H}\rangle+|B^{0}_{L}\rangle;~
|\bar
B^{0}_{d}\rangle=\frac{1}{2N}~\frac{p}{q}~(|B^{0}_{H}\rangle-|B^{0}_{L}\rangle
),
\end{equation}
Thus, in the strong interaction, {\em coherent superpositions}  of
states of $B^{0}_{H}$ and $B^{0}_{L}$ mesons, particles with
 definite masses and widths, are produced.

Let us compare relations  (\ref{85}) with neutrino relations  (\ref{31}) and (\ref{32}). 
Flavor neutrinos and antineutrinos $\nu_{l}$ and $\bar \nu_{l}$ are produced in weak processes
in which at neutrino energies 
many order of magnitudes larger than neutrino masses   lepton flavor numbers 
$L_{e}$, $L_{\mu}$ and $L_{\tau}$ are effectively  conserved.
The states 
of these neutrinos are coherent superpositions of the states of neutrinos with 
definite masses.  Flavor 
$B_{d}^{0}$ and $\bar B_{d}^{0}$ mesons are produced because quark flavor is conserved in 
strong interaction.  Their states are coherent superposition (\ref{85}).

There exists also an  important difference between mixing of states of
neutral mesons and mixing of neutrino states. Neutrino mixing is
determined by the PMNS mixing matrix. Even in the case of the two
neutrinos mixing angle is an arbitrary parameter. In the case of
neutral mesons mixing is maximal (independently on the values of CKM
mixing angles). This is connected with CPT invariance of the strong
interaction and the fact that $B^{0}_{d}$ and $\bar
B^{0}_{d}$ are transformed into each other under CP transformation.
CPT invariance does not put any constraints on neutrino mixing
angles.

We will consider now time evolution of the states of  of  $B^{0}_{d}$ and
$\bar B^{0}_{d}$ mesons.  From (\ref{80}),  (\ref{82})  and  (\ref{85}) we find
\begin{equation}\label{86}
|B^{0}_{d}\rangle_{t}=\frac{1}{2N}~(|B^{0}_{H}\rangle~e^{-i\lambda_{H}t}+
|\bar B^{0}_{L}\rangle~e^{-i\lambda_{L}t})
\end{equation}
and
\begin{equation}\label{87}
|\bar
B^{0}_{d}\rangle_{t}=\frac{1}{2N}~\frac{p}{q}~(|B^{0}_{H}\rangle~e^{-i\lambda_{H}t}-
|\bar B^{0}_{L}\rangle~e^{-i\lambda_{L}t}),
\end{equation}
where $t$ is the proper time.

Neutral $B$ mesons are detected through the observation of the
decays of $B_{d}^{0}$ and $\bar B_{d}^{0}$, which are determined by 
transitions $\bar b \to\bar c (\bar u)+W^{+} $ and
$b \to c ( u)+W^{-} $. For lepton decays we have
$$B_{d}^{0}\to l^{+}+\nu_{l}+X ;~~~\bar B_{d}^{0}\to l^{-}+ \bar\nu_{l}+X.$$
 Thus, the
sign of the charged lepton determines the type of the neutral
$B$-meson. From (\ref{86}) and (\ref{87}) we find
\begin{equation}\label{88}
|B^{0}_{d}\rangle_{t}=g_{+}(t)~|B^{0}_{d}\rangle
+\frac{q}{p}~g_{-}(t)~|\bar B^{0}\rangle
\end{equation}
and
\begin{equation}\label{89}
|\bar B^{0}_{d}\rangle_{t}=g_{-}(t)~\frac{p}{q}~|B^{0}_{d}\rangle
+g_{+}(t)~|\bar B^{0}\rangle.
\end{equation}
Here
\begin{equation}\label{90}
g_{\pm}(t)=\frac{1}{2}~(e^{-i\lambda_{H}t}\pm e^{-i\lambda_{L}t})
\end{equation}
The equations (\ref{88}) and (\ref{89}) are analogous to the
equations (\ref{45}) and (\ref{46}) in the neutrino case. For the
transition probabilities  we find the following expression
\begin{equation}\label{91}
P(B_{d}^{0}\to \bar B_{d}^{0})= P(\bar B_{d}^{0}\to B_{d}^{0})=
\frac{1}{2}~e^{-\Gamma t}(1-\cos\Delta m_{B}~t),
\end{equation}
where
\begin{equation}\label{92}
\Delta m_{B} = m_{H}-m_{L}
\end{equation}
is the difference of masses of $B^{0}_{H}$ and $B^{0}_{L}$ mesons.
We took into account in (\ref{91}) 
that $\Gamma_{H}\simeq \Gamma_{L}=\Gamma$ and $|\frac{p}{q}|\simeq 1$
 (see \cite{CPreviews})

The probability of $B^{0}$ ($\bar B^{0}$) to survive is given by the
expression
\begin{equation}\label{93}
P(B_{d}^{0}\to B_{d}^{0})= P(\bar B_{d}^{0}\to \bar B_{d}^{0})=
\frac{1}{2}~e^{-\Gamma t}(1+\cos\Delta m_{B}~t).
\end{equation}
The equations (\ref{91}) and (\ref{93}) are analogous to equations
(\ref{67}) and (\ref{68}) in the neutrino case.

The main differences between neutrino oscillations and
$B_{d}^{0}\leftrightarrows\bar B_{d}^{0}$ oscillations are  the
following
\begin{itemize}
\item In the $B$-mesons case mixing is maximal. In the
neutrino case mixing angles are parameters. Investigation
of neutrino oscillations is the only possible source of information
about the elements of the PMNS mixing matrix. Modulus of elements of
the quark CKM mixing matrix can be determined from the investigation
of different weak decays and neutrino reactions. Information about
CP phase can be inferred from the measurement of CP-odd asymmetry in
$B_{d}^{0}(\bar B_{d}^{0})\to J/\psi +K_{S}$ and other
decays (see \cite{bfactories}).

\item Neutral $B$-mesons are unstable particles. Life-time is given
by $\simeq \frac{1}{\Gamma}$ and time-energy uncertainty relation
\begin{equation}\label{94}
\Delta m_{B}~\frac{1}{\Gamma}\geq 1
\end{equation}
is the condition of the observation of
$B_{d}^{0}\leftrightarrows\bar B_{d}^{0}$ oscillations. 

Neutrinos
are stable particles. The time $t\simeq L $ in the time-energy uncertainty
relation (\ref{49}) is determined by  experimental conditions
(intensity of a neutrino beam, size of a detector, background
conditions etc)
\end{itemize}
The formalism of $B_{d}^{0}\leftrightarrows\bar B_{d}^{0}$
oscillations, which we shortly discussed here  (and similar formalism of
$K^{0}\leftrightarrows\bar K^{0}$ oscillations) is based on the
evolution equation (\ref{80}), mixed flavor states,  non stationary states 
$|B^{0}_{d}\rangle_{t}$ and $|\bar B^{0}_{d}\rangle_{t}$, given by  (\ref{88}) and (\ref{89}). 
 This formalism describes a lot of
precise experimental data. From our point of view correct formalism
of neutrino oscillations also must be based on the  Schrodinger evolution equation, mixed 
flavor neutrino states and time-energy uncertainty
relation.

\section{Conclusion}

Observation of neutrino oscillations in SK, SNO, KamLAND and other
neutrino experiments \cite{SK,SNO,Kamland,K2K,Cl,Gallex,Sage,SKsol} is 
an important recent discovery in the particle physics. 
Investigation of this new phenomenon allowed to determine such 
fundamental parameters as 
neutrino mass squared differences and neutrino mixing angles. It is a common opinion that 
generation of neutrino masses, which are many orders of magnitude smaller than masses of leptons and quarks, require a new physics and a new beyond the Standard Model mechanism of the mass generation.

Taking into account importance of neutrino oscillations for particle physics we think that the 
understanding of the physical basis of this new phenomenon is an important issue.
In literature exist different interpretations of the basics of neutrino oscillations (see \cite{Carlo, Kayser}).
 We present here
the following point of view:

\begin{itemize}
\item Neutrino masses are many orders of magnitude smaller
than energies of neutrinos in reactor, solar, atmospheric and
accelerator neutrino experiments. This means that  neutrino masses can be neglected in
matrix elements of neutrino-production and neutrino-detection
processes. Therefore in CC weak processes together with $l^{+}$
($l^{-}$) flavor left-handed neutrinos $\nu_{l}$ ( flavor
right-handed antineutrinos $\bar\nu_{l}$) are  produced which states  are described by coherent superpositions
of states of neutrinos with definite masses. The possibility to
neglect neutrino masses in production and detection processes signifies
that in these processes flavor lepton numbers are effectively
conserved.
\item
Flavor neutrinos and antineutrinos at time $t$ after
production are described by non
stationary states (coherent superposition of stationary states). For
such states time-energy uncertainty relation is valid. 
This means that neutrino oscillations with finite oscillation time 
require uncertainty of energy.
\item
The presented here non stationary formalism of neutrino oscillations 
is the same as well known formalism of $K^{0}\leftrightarrows\bar
K^{0}$, $B_{d}^{0}\leftrightarrows\bar B_{d}^{0}$ etc oscillations, which is 
confirmed by numerous high precision experiments.

\end{itemize}
{\bf Acknowledgements.} M.M. would like to thank 
the Elementary Particle Physics Sector of SISSA, 
where a part of this work was done, for the kind hospitality.
Both authors would like to thank  S. Petcov and C. Giunti
for friutful discussions. S.M.B. acknowledges the support of the MIUP  Italian
program ``Rientro dei Cervelli''. 
This work was also supported in part by the Italian MIUR 
program ``Fisica Astroparticellare'' (M.M.) ,
by the Bulgarian National Science Fund
under the contract Ph-09-05 (M.M.) and by the 
the European Network of Theoretical Astroparticle Physics 
ILIAS/N6 under the contract RII3-CT-2004-506222
(S.M.B.).

\end{document}